\documentclass[prl,aps,a4,epsf,8pt,twocolumn,showpacs]{revtex4}

\usepackage{amsmath}
\usepackage{amssymb}
\usepackage{graphicx}
\usepackage[usenames]{color}

\begin{document}

\newcommand{\blue}[1]{\textcolor{blue}{#1}}

\newcommand{\beq}{\begin{equation}}
\newcommand{\eeq}{\end{equation}}
\newcommand{\beqa}{\begin{eqnarray}}
\newcommand{\eeqa}{\end{eqnarray}}
\newcommand{\bmat}{\begin{displaymath}}
\newcommand{\emat}{\end{displaymath}}

\newcommand{\eq}[1]{Eq.~(\ref{#1})}

\newcommand{\lan}{\langle}
\newcommand{\ran}{\rangle}

\newcommand{\tav}[1]{\left\lan #1 \right\ran}

\title{Rigidity of thermalized soft repulsive spheres around the jamming point}

\author{Satoshi Okamura and Hajime Yoshino}

\affiliation{Department of Earth and Space Science, Faculty of Science,
 Osaka University, Toyonaka 560-0043, Japan}

\begin{abstract}
We study the effect of thermalization on the 
rigidity of a randomly packed soft repulsive sphere system 
around the jamming point by analyzing the shear-modulus
using the cloned liquid theory with
the 1 step replica symmetry breaking ansatz and molecular dynamics simulations.
Contrarily to the usual harmonic picture for solids,
we found that the thermalized jamming system 
is anomalously softer than at zero temperature 
such that the shear-modulus becomes as small as the pressure
down to vanishingly low temperatures. 
\end{abstract}


\date{\today}
\maketitle

Randomly packed repulsive foams, colloids, emulsions and granular particles
exhibit anomalous solid states.
At high densities liquid like spontaneous structural rearrangements of the positions
of the particles become almost impossible much as deeply supercooled molecular glass forming liquids. 
However the contacts between the particles exhibit surprisingly rich dynamic characters
around the jamming point located deep in the glassy phase as manifested 
in static properties
\cite{Mason-Bibette-Weitz-1995,Durian-1995,Mason-et-al-97,Nagel-group,Wyart-2006,Brito-Wyart-2006,Brito-Wyart-2009,Ellenbroek-et-al-2009,Otsuki-Hayawaka-2012,NYU-exp-group-2013},
linear and non-linear rheology \cite{Mason-Bibette-Weitz-1995,Mason-et-al-97,Durian-1995,Olsson-Teitel,Hatano,Ikeda-Berthier-Sollich-2012} and relaxations \cite{Coulais-Bertinger-Dauchot-2012,Ikeda-Berthier-Biroli-2012}. 


An important common ingredient in this class of systems 
is that the particles are interacting with each other through repulsive
contact potentials with some definite cut-off at particle scales. 
Quite interestingly recent  works \cite{Schreck-et-al-2011,Wyart-2012,Lener-et-al-2013} suggest that the solid states of such systems are only marginally stable against unharmonic, plastic deformations such that opening of a weak contact 
can trigger an avalanche like deformation. 
This is contrarily to the usual view that harmonic descriptions work well 
in solids because defects would require finite creation energies so that they 
become suppressed at low enough temperatures. 

On the other hand, physically we would still expect that 
these glassy systems, even if they are  not 
strictly harmonic, retain some sort of macroscopic 
solidity at least at low enough temperatures. 
A natural quantity to quantify the solidity is the shear-modulus or the
{\it rigidity} $\mu$, which is
finite only in solids. Indeed a set of experiments on an emulsion system
\cite{Mason-Bibette-Weitz-1995,Mason-et-al-97}  
done at the room temperature which amounts to a rescaled temperature 
$T \sim O(10^{-5})$ have made a striking
observation that the system is anomalously soft such that 
both the entropic rigidity at volume fractions $\varphi < \varphi_{\rm J}$
smaller than the jamming density $\varphi_{\rm J}$
and more mechanical rigidity at $\varphi > \varphi_{\rm J}$
behaves just as the pressure $P$,
i.~e $\mu \sim P$ while the harmonic picture predicts $\mu \gg P$ for both regimes \cite{Brito-Wyart-2006,Brito-Wyart-2009,Ellenbroek-et-al-2009}.

The purpose of this paper is to study a model system
which mimic the emulsion system \cite{Princen-1983} 
and examine the rigidity at low
but finite temperatures by theoretical and numerical approaches. 
For the theoretical approach we employ the cloned liquid method
\cite{mezard-parisi-1999,parisi-zamponi-2010,Berthier-Jacquin-Zamponi-2011}
which is a 1st principle method based on the liquid theory
and the replica method aimed to analyze static properties
of glasses and jamming systems within the scenario of the
random first order transition (RFOT) theory 
(see \cite{RFOT-review} for a review).
To compute the rigidity in this framework we follow 
the scheme of \cite{YM2010,Y2012} and extend it
to the jamming system.
We also performed simulations of the
shear-stress relaxation to obtain the thermalized rigidity.
Our main results is that the thermalized rigidity behaves
just as the pressure much as in the experiment
but in contradiction to the harmonic picture.
We argue that our result 
suggests some gap-less spectrum of plastic excitations
between different inherent structures (energy minima) 
within a meta-basin (metastable state) \cite{Doliwa-Heuer-2003}. 

{\bf Model} We study a system of $N$ particles 
($i=1,2,\ldots,N$) of diameter $a=1$ in the $3$-dimensional space with
volume $V$, which interact with each other through a soft repulsive
contact potential given below. 
The crucial parameter is the volume fraction $\varphi=(\pi/6)a^{3}\rho$ 
which is related to the number density $\rho=N/V$.
We denote the positions of the particles as ${\bf r}_{i}$
and represents a pair of particles $i$ and $j$ as $\langle i j \rangle$.
The potential energy is given by
$
U = \sum_{\langle i j \rangle} \phi(r_{ij})$
where $r_{ij}=|{\bf r}_{i}-{\bf r}_{j}|$.
We consider a soft repulsive contact potential of
the form,
$
\phi(r)=\epsilon(1-r/a)^{2}\theta(1-r/a)$,
where $\theta(r)$ is the step function. This is considered as an
approximate model potential for emulsions \cite{Princen-1983,NYU-exp-group-2013}.
We denote a rescaled temperature $k_{\rm B}T/\epsilon$, where $k_{\rm
B}$ is the Boltzmann's constant, simply as $T$ below.

{\bf Cloned liquid approach} In the cloned liquid approach we consider a
system of $m$ replicas, each of which is a system of $N$ 
particles.
In order to study the glassy phase we employ
the 1 step replica symmetry breaking (RSB) ansatz
which is known to capture the 
phenomenology of glasses \cite{RFOT-review}.
The 1-RSB ansatz amounts to consider a sort of
'molecular' liquid \cite{mezard-parisi-1999}.
The coordinates 
of the particles ${\bf r}^{a}_{i}$ ($a=1,\ldots,m$; $i=1,2,\ldots,N$)
is decomposed  as,
${\bf r}^{a}_{i}={\bf R}_{i}+{\bf u}_{i}^{a}$
with
${\bf R}_{i}\equiv \frac{1}{m}\sum_{a=1}^{m}{\bf r}^{a}_{i}$
where ${\bf u}_{i}^{a}$ stands for fluctuation 
of the particle belonging to the $a$-th replica
with respect to the center of mass ${\bf R}_{i}$ of a 'molecule'. 
Physically the size of the molecule which we denote as $A$ 
represents the size of cage in the metastable states.
The fluctuations within the molecules (cages)
are assumed to obey the Gaussian statistics 
with the mean $0$ and variance 
 \cite{mezard-parisi-1999}, 
$\langle ({\bf u}^{a}_{i})^{\mu}({\bf u}^{b}_{j})^{\nu} \rangle_{\rm
cage} = A(\delta_{ab}-\frac{1}{m})\delta_{\mu \nu}\delta_{ij}$.
Here $\mu$ (and $\nu$)
represents a component of 3-dimensional vectors $\mu=x,y,z$.
These parameters $A$ and $m$ 
are optimized to minimize the free-energy $F_{m}/m$.

In order to study static
response to shear $(x_{ij} \to x_{ij}+\gamma z_{ij},
y_{ij} \to y_{ij}, z_{ij} \to z_{ij})$
of the replicated system, the free-energy $F_{m}$ of the whole system
would be expanded in power series of the shear-strain
$\gamma_{a}$ $(a=1,2,\ldots,m)$ as  \cite{YM2010,Y2012}, 
$F_{m}(\{\gamma\})/V=F_{m}(\{0\})/V+\sum_{a=1}^{m}\sigma_{a}\gamma_{a}  +(1/2)\sum_{a,b=1}^{m}
\mu_{ab}\gamma_{a}\gamma_{b} +\ldots$.
where $\sigma_{a}$ is the shear-stress 
of $a$-th replica, i.~e. 
$\sigma_{a}=(1/V)\sum_{\langle i,j\rangle} \langle \sigma({\bf r}^{a}_{ij})\rangle$ 
with $\sigma({\bf r}) \equiv \hat{z}\hat{x}r \phi'(r)$.
Here we introduced short-hand notations like
$\hat{x} \equiv x/r$ with $r=|{\bf r}|$. 
The shear-modulus or rigidity (matrix) $\mu_{ab}$ can be expressed by a fluctuation formula,
\begin{eqnarray}
&&  
\beta \mu_{ab} = \frac{1}{V} \left [  \sum_{\langle kl\rangle} \langle \beta\sigma({\bf r}^{a}_{kl})\rangle
\langle \beta \sigma_{b}({\bf r}^{b}_{kl})\rangle  \right.  
\label{eq-mu-ab} \\
&&  
\left. -
\sum_{\langle mn \rangle \neq \langle kl \rangle}
(\langle \beta \sigma({\bf r}^{a}_{kl}) \beta \sigma({\bf r}^{b}_{mn})\rangle 
\right. 
\left. 
-\langle \beta \sigma({\bf r}^{\bf a}_{kl})\rangle\langle \beta\sigma({\bf r}^{b}_{mn})\rangle)
\right] \nonumber
\end{eqnarray}
where $\beta=1/T$ and $\langle \ldots \rangle$ represents a thermal average.
which is the local shear-stress associate with a pair of particles.
The expression \eq{eq-mu-ab}
is derived by simplifying the one given in 
\cite{Farago-Kantor-2000} assuming rotational symmetry, which is valid
in the present system,  and generalizing the formula to the replicated system.
It is more useful than other equivalent expressions 
involving the 2nd derivative of the potential 
which can be problematic for the contact potential systems.

Within the 1-RSB ansatz we 
expect that the rigidity matrix
$\mu_{ab}$ takes the generic form \cite{YM2010,Y2012},
$\mu_{ab}=\hat{\mu}\left(\delta_{ab}-\frac{1}{m} \right)$.
Here $\hat{\mu}=-m \mu_{a \neq b}$ can be regarded as the rigidity
of metastable states which is our main concern in the present paper.

The thermodynamic properties of
the present system were studied in detail by the cloned liquid approach 
at the level of 1-RSB in \cite{Berthier-Jacquin-Zamponi-2011}.
They are the same as those of the hard-sphere systems  \cite{parisi-zamponi-2010}
in the low temperature limit at $\varphi < \varphi_{\rm GCP}$.
Here $\varphi_{\rm GCP}$ is the so called glass close packing (GCP) density 
which is an {\it ideal} jamming density where the reduced pressure 
$\beta P/\rho$ of the {\it equilibrium} hard-sphere glass state diverges.

We can assume that the cage size $A$ is very small 
deep in the glassy phase.
First we note that if $A=0$, $\mu_{ab}=0$
since the cloned liquid as a whole is just a liquid.
Next we consider contribution from the fluctuations inside the cages.
Focusing on the 2nd term on the r.h.s. of \eq{eq-mu-ab} we notice that 
it can be separated into three-point 
terms ${\sum}_{i < j < k} \ldots$ and four-point terms ${\sum}_{i< j < k < l} \ldots$.
The 1st non-vanishing contribution to $\hat{\mu}$ at order $O(A)$ 
comes from the three-point terms,
\begin{widetext}
\begin{eqnarray}
&&\beta\hat{\mu} =  
\left. 
-A\frac{\rho}{V}
\int d^{d}r_{0} d^{d}r_{1} d^{d}r_{2}
\left( \nabla_{0}^{a} \cdot \nabla_{0}^{b}\right)
\beta\sigma({\bf r}_{01}^{a})
\beta\sigma({\bf r}_{02}^{b})
g_{3}(r^{a}_{01},r^{b}_{02},r_{12};T^{*})
\right |_{{\bf r_{01}^{a}}={\bf r}_{01},{\bf r_{02}^{b}}={\bf
r}_{02}} + \ldots \qquad (a \neq b) \\
\label{eq-hat-mu-hardsphere}
&& \simeq
-\frac{A}{m^{2}} \rho \int dr_{1} dr_{2} d\Omega_{1}d\Omega_{2}
      \nabla_{1}\left[ \hat{x}_{1}\hat{z}_{1} r_{1} \Delta(r_{1};T^{*}) y(r_{1};T^{*})\right]
\cdot \nabla_{2}\left[\hat{x}_{2}\hat{z}_{2} r_{2} \Delta(r_{2};T^{*}) y(r_{2};T^{*}) \right]y(r_{12};T^{*})e^{-\beta^* \phi(r_{12})}
\label{eq-hat-mu-hardsphere-kirkwood}
\end{eqnarray}
\end{widetext}
where we introduced a short-hand notation $T^{*} \equiv T/m$ 
and $\beta^{*}=1/T^{*}$. 
In the 2nd equation we approximated the three-point correlation 
function $g_{3}({\bf r}_{1},{\bf r}_{2})$ by the Kirkwood approximation
in order to make an analytical progresses,
$
g_{3}(r_{1},r_{2},r_{12}) \simeq g(r_{1})g(r_{2})g(r_{12})
$
where $g(r)$ is the radial distribution function.
We switched to the polar coordinates
with radial variables $r_{1}$, $r_{2}$ and solid angles $\Omega_{1}$, $\Omega_{2}$.
We also introduced the cavity function
$
 y(r;T)  \equiv e^{\beta \phi(r)} g(r;T)
$
and 
a function
$
\Delta(r;T)  \equiv \frac{d}{dr}e^{-\beta \phi(r)}
$
which becomes a delta function in the $T \to 0$ limit enabling analytical computations.

We also have to take into account the renormalization of the interaction between 'molecules' due to
fluctuations inside the molecules \cite{mezard-parisi-1999,parisi-zamponi-2010,Berthier-Jacquin-Zamponi-2011}.
It amounts to replace the original potential by a renormalized one 
$m\phi(r) \to \varphi_{\rm eff}(r)$ in the above computation. 
For the present system we find
$e^{-\beta \varphi_{\rm eff}(r)} \simeq \theta(r-a) +\sqrt{\pi A/m}\delta(r-a)+ \ldots$ 
around the jamming point $\varphi=\varphi_{\rm GCP}$ using the results of
\cite{Berthier-Jacquin-Zamponi-2011} 
and assuming that $A/m$ is small. Finally we obtain,
\beq
\hat{\mu} =  \frac{T}{m}
\frac{A}{m} 
\frac{6\varphi}{\pi a^{3}} y(a;T^{*},\varphi)^{3}
\left[\alpha_{1}-\alpha_{2}\sqrt{\frac{A}{m}} + \ldots \right]
\label{eq-hat-mu-hardsphere-2}
\eeq
with $\alpha_{1}=(113/120)\pi^{2}$ and
$\alpha_{2}=(376709/22050)\pi^{2}$.

The above expression \eq{eq-hat-mu-hardsphere-2} implies that the scaling property
of the rigidity $\hat{\mu}$ around the jamming point $\varphi=\varphi_{\rm GCP}$
is dominated by that of the parameter $m$.
The latter is known to behave as  \cite{Berthier-Jacquin-Zamponi-2011}, 
$
m(T,\varphi)=\sqrt{T}\tilde{m}\left((\varphi-\varphi_{\rm GCP})/\sqrt{T}\right)$
with $\tilde{m}(x) \simeq c_{1} |x|$ for $x < 0$ and 
$\tilde{m}(x) \simeq c'_{1}/x$ for $x > 0$.
Now we find the rigidity behaves around $\varphi=\varphi_{\rm GCP}$ as,
\beq
\lim_{T \to 0} \hat{\mu} =c_{+}(\varphi-\varphi_{\rm GCP})
\qquad \varphi > \varphi_{\rm GCP}.
\label{eq-scaling-hat-mu-above}
\eeq
and
\beq
\lim_{T \to 0} \beta \hat{\mu} =\frac{c_{-}}{\varphi_{\rm GCP}-\varphi}  
\qquad \varphi < \varphi_{\rm GCP}
\label{eq-scaling-hat-mu-below}
\eeq
which can be regarded as an entropic rigidity.
Here the numerical prefactor are $c_{-}=cc_{1}$ and $c_{+}=cc'_{1}$
where $c=(6\varphi_{\rm GCP}/\pi)\alpha_{\rm GCP}y(1;0,\varphi_{\rm
GCP})^{3}(a_{1}-a_{2}\sqrt{\alpha_{\rm GCP}})$ with $\alpha_{\rm GCP}$
being the optimized value of $A/m$ at $\varphi=\varphi_{\rm GCP}$. 
Using the values of  $\alpha_{\rm GCP}$, 
$\varphi_{\rm GCP}$, $y(\varphi_{\rm GCP})$, $c$ and $c'$ given in \cite{Berthier-Jacquin-Zamponi-2011}, we find 
$c_{+} \simeq 0.1239496$ and $c_{-} \simeq 0.694315$.

A remarkable feature is that the behaviour of the rigidity 
$\hat{\mu}$ found above is exactly the same as that of
the pressure $P$  \cite{parisi-zamponi-2010,Berthier-Jacquin-Zamponi-2011,Otsuki-Hayawaka-2012}.
Quite interestingly this is consistent with the result of the 
the experiment on the emulsion \cite{Mason-Bibette-Weitz-1995,Mason-et-al-97}  
and the MD simulations at finite temperatures \cite{Okamura-2012}. 

However the result \eq{eq-scaling-hat-mu-above} apparently contradicts
with the behaviour at $T=0$ \cite{Durian-1995,Nagel-group,Wyart-2006}, i.~e. $\mu_{\rm harmonic} \propto \sqrt{\varphi-\varphi_{\rm J}}$ for $\varphi > \varphi_{\rm J}$. If the solid state is harmonic, this scaling must hold at low temperatures. 
Remarkably an effective harmonic solid picture has been developed 
also below the jamming density $\varphi < \varphi_{\rm J}$ \cite{Brito-Wyart-2006,Brito-Wyart-2009}.
The latter predicts
$\mu_{\rm harmonic} \propto T/(\varphi_{\rm J}-\varphi)^{3/2}$ for $\varphi <
\varphi_{\rm J}$,
which is again different from \eq{eq-scaling-hat-mu-below}. 
Thus both below and above the jamming density $\mu_{\rm hamonic} \gg P$
while we find the thermalized rigidity as $\mu \sim P$.
Concerning the behaviour of pressure $P$ itself, 
there seems to be no contradictions 
among the emulsion experiment \cite{Mason-Bibette-Weitz-1995,Mason-et-al-97,NYU-exp-group-2013}, 1-RSB theory \cite{parisi-zamponi-2010,Berthier-Jacquin-Zamponi-2011}, simulations done at zero \cite{Durian-1995,Nagel-group} and finite temperatures \cite{Otsuki-Hayawaka-2012}.
In order to solve the paradox, we next examine 
shear-stress relaxation by MD simulations.

\begin{figure}[t]
  \begin{center}
   \resizebox{0.4\textwidth}{!}{\includegraphics{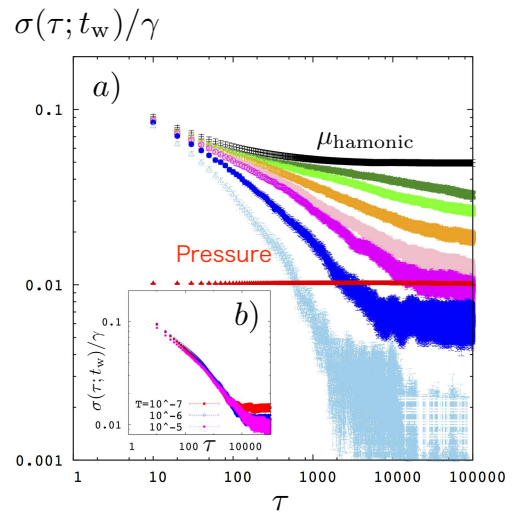}}
  \caption{\small Relaxation of the shear-stress in the system
with $\varphi=0.67$ ($\varphi_{\rm J} \sim 0.648$)
at $T=10^{-5}$. The initial temperature is 
$T_{\rm i}=10^{-2}$ where the system is in the liquid state.
The number of particles used is $N=800$.
Comparing with some data of $N=2400$ we found no appreciable finite size effects
within our time window. The strength of the shear-strain is
$\gamma=2.5 \times 10^{-3}$. Comparing with some data of 
$\gamma=2.5 \times 10^{-4}$ we confirmed that the linear response holds.
The average over initial configurations is taken over $4096$ samples.
{\bf Panel a)}: data are obtained at $T=10^{-5}$ after waiting times
$t_{\rm w}=3 \cdot 10^{2},10^{3},3 \cdot 10^{3}, 5 \cdot 10^{3},10^{4},3 \cdot 10^{4},10^{5}$ from the bottom to the upper curves. The top curve (indicated as $\mu_{\rm harmonic}$) is the data obtained at $T=0$ after aging for $t_{\rm w}=10^{5}$ at $T=10^{-5}$.
The data of the pressure $P$ ($T=10^{-5}$, $t_{\rm w}=10^{5}$)
is also shown for a comparison.
{\bf Panel b)}: data at different temperatures $T=10^{-5},10^{-6},10^{-7}$
with $t_{\rm w}=3\cdot 10^{3}$ are shown for a comparison.
}
\label{fig-stress-relaxation}
  \end{center}
\end{figure}

{\bf MD simulations of shear-stress relaxation}
A natural protocol to probe the rigidity of glassy states is 
to observe the {\it plateau} modulus which appears in the shear-stress relaxation processes.
To study the relaxation deep in the glassy regime systematically, 
we employ the standard protocol of aging experiments \cite{mckenna-group}:
(1) thermalize the system at a high temperature $T_{\rm i}$
(2) quench the temperature down to the working temperature $T$ and
let the system relax for a waiting time $t_{\rm w}$
(3) put a small shear-strain $\gamma$ on the system
(4) measure the shear-stress $\sigma(\tau)$ as a function of the 
time $\tau$ elapsed afterwards.

A bi-disperse system of the ratio of radii $1.4$ is employed.
The relaxation is simulated by solving
the under-damped Langevin dynamics using
a velocity Verlet algorithm with an integration step $h=0.01$.
The unit of time $t_{\rm micro}=1$ is the relaxation time of the non-interacting system.  The shear-strain $\gamma$ is applied by an affine transformation and the Lee-Edwards boundary conditions.

In Fig.~\ref{fig-stress-relaxation} a), b) we show a representative set of data
of the stress relaxation.
A notable feature is that there are two kinds of plateaus: the higher one at shorter $\tau$ and lower one at longer $\tau$. The relaxation between the former to the latter strongly depend on the waiting time $t_{\rm w}$. 

We interpret the two plateaus as the following. We regard the higher plateau as the rigidity $\mu_{\rm harmonic}$ of inherent structures (IS) or energy minima, which is
essentially the same as rigidity at zero temperature \cite{Durian-1995,Nagel-group}.
For sufficiently short time scales $\tau$, the dynamics should be harmonic so that 
$\mu_{\rm harmonic}$ only takes into account harmonic thermal fluctuations around ISs.
Indeed we observe that the initial part of the dynamics depend little on
the temperature. The plateau value $\mu_{\rm harmonic} \simeq 0.047$ agrees well with the value obtained directly at zero temperature \cite{Nagel-group}.

On the other hand the other plateau is much lower than $\mu_{\rm harmonic}$ and lies at the level
of the pressure $P$. This would be interpreted as the rigidity of metabasins (MB) which is
a union of ISs \cite{Doliwa-Heuer-2003}. The strong waiting time
dependence implies that the system is searching for ISs with lower
free-energy within a MB. 
Then we may interpret the rigidity \eq{eq-scaling-hat-mu-below}
and \eq{eq-scaling-hat-mu-above} found at the level of 1-RSB as the
rigidity of the MBs rather than ISs. 

\begin{figure}[t]
  \begin{center}
   \resizebox{0.4\textwidth}{!}{\includegraphics{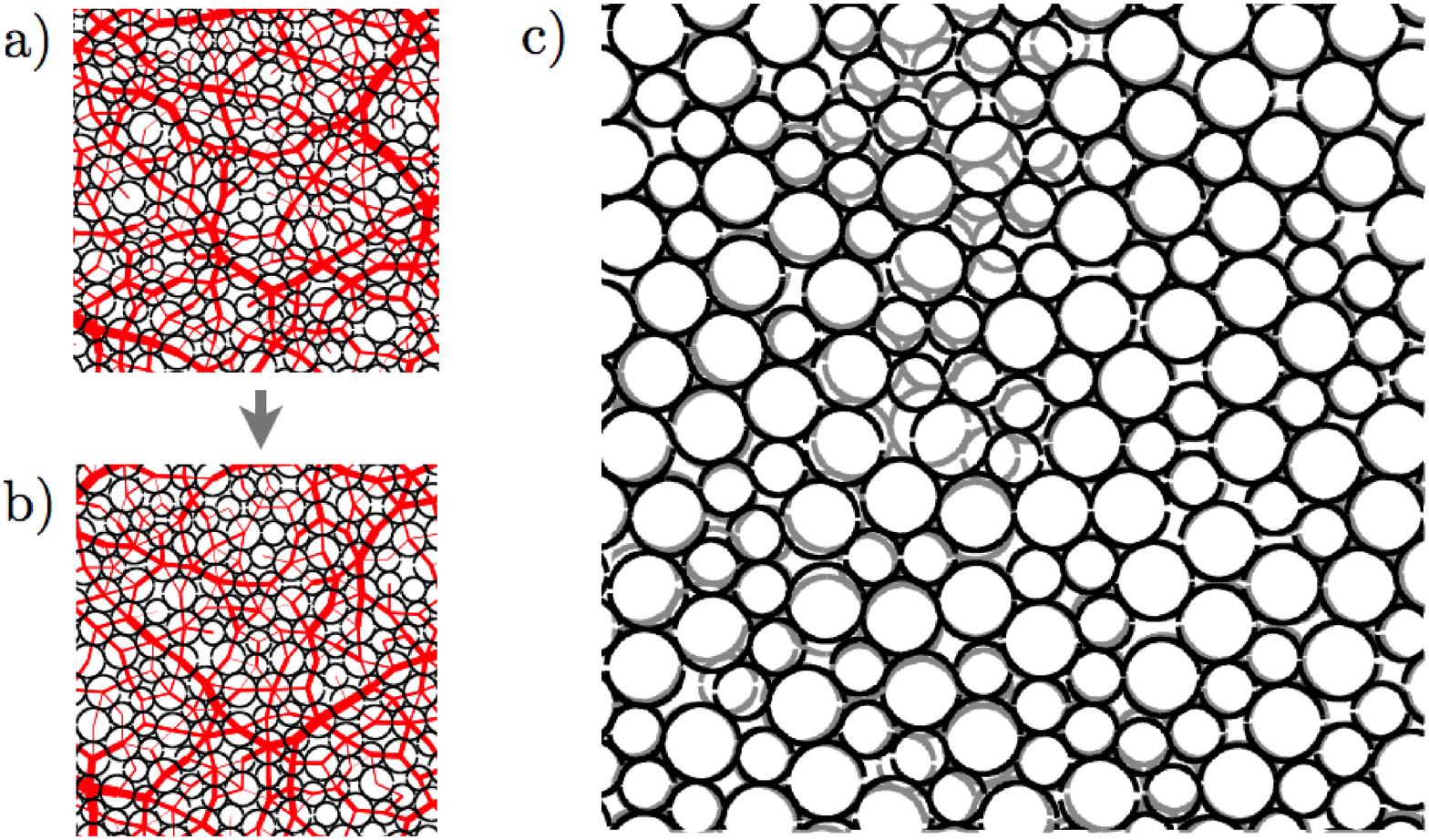}}
  \end{center}
 \vspace*{-5mm}
  \caption{\small Snapshots of a plastic event
during stress-relaxation. Here we used a 2-dimensional version of the model 
at  $\varphi=0.85$ ($\varphi_{\rm J}\sim 0.84$).
The system is initially perturbed by $\gamma=0.05$
and let to relax at $T=10^{-6}$.
The configuration of particles are represented by the circles and that of the 
contact forces $f_{ij}=-d\phi(r_{ij})/dr_{ij}$ are
represented by bonds whose thickness is proportional to $f_{ij}$.
The panels a) and b) show the snapshots before/after a plastic event 
(whose duration is about $10^{4}$).
In panel c) the configuration of the particles before/after the event are 
shown by lighter/darker colors.}
\label{fig-snapshot}
\end{figure}

{\bf Discussions} 
The transitions between the ISs inside a MB necessarily involve plastic processes. 
Indeed we can see avalanche like processes as shown in Fig.~\ref{fig-snapshot} 
at the time scale corresponding to the shear-stress relaxation between the two plateaus. 
We speculate that they are the floppy modes 
which are released by opening contacts \cite{Wyart-2012}. 
Presumably they are extended 
over the so called isostatic length scale \cite{Wyart-2006}.
It is natural to expect that 
the distribution of 
the excitation energies of the floppy modes 
is gap-less so that there is a fraction of active floppy modes even at vanishingly low temperatures.
The floppy modes do not change the distances between the particles in contact 
but 
can {\it rotate} the contacts.
This scenario naturally explains why there is a significant shear-stress relaxation
while the pressure remains almost constant in time as shown in Fig.~\ref{fig-stress-relaxation}.
From the theoretical point of view, it would be very interesting to try to understand
these features in terms of the  $1+\infty$ RSB scenario \cite{Kurchan-Parisi-Urbani-Zaompni-2013}.

To conclude, we found that the 
rigidity of the repulsive jamming system survives at finite temperatures 
but in a peculiar way:
the shear-modulus becomes substantially smaller than at zero temperature
even at vanishingly low temperatures, contrary to usual crystals and glasses.

We thank Francesco Zamponi, Atsushi Ikeda, Ludovic Berthier, Michio Otsuki, Olivier Dauchot, Guerra Rodrigo and Mathieu Wyart for useful discussions.
This work is supported by  Grant-in-Aid for Scientific Research (C) (50335337)
and JPS Core-to-Core Program ``Non-equilibrium dynamics of soft matter
and informations''. The numerical simulation has been done using the facilities of the Supercomputer Center, the Institute for Solid State Physics, the University of Tokyo.

\end{document}